\documentclass[
 aip,
 amsmath,amssymb,
 reprint,%
]{revtex4-2}
\usepackage[normalem]{ulem}
\usepackage{graphicx}
\usepackage{dcolumn}
\usepackage{bm}
\usepackage{xcolor}
\usepackage[utf8]{inputenc}
\usepackage[T1]{fontenc}
\usepackage{mathptmx}
\usepackage{etoolbox}
\usepackage{comment}
\usepackage[english]{babel}
\makeatletter
\def\@email#1#2{%
 \endgroup
 \patchcmd{\titleblock@produce}
  {\frontmatter@RRAPformat}
  {\frontmatter@RRAPformat{\produce@RRAP{*#1\href{mailto:#2}{#2}}}\frontmatter@RRAPformat}
  {}{}
}%
\makeatother
\begin{document}

\preprint{AIP/123-QED}

\title[]{Dynamic coupling and spin-wave dispersions in a magnetic hybrid system made of an artificial spin-ice structure and an extended NiFe  underlayer}

\author{R. Negrello}
\author{F. Montoncello}%
\affiliation{ 
Dipartimento di Fisica e Scienze della Terra, Università di Ferrara, Ferrara, Italy
}%

\author{M. T. Kaffash}
\affiliation{Department of Physics and Astronomy, University of Delaware, Newark, Delaware 19716, USA
}%
\author{M. B. Jungfleisch}%
 \email{mbj@udel.edu}
\affiliation{Department of Physics and Astronomy, University of Delaware, Newark, Delaware 19716, USA}

\author{G. Gubbiotti}%
\affiliation{Istituto Officina dei Materiali del Consiglio Nazionale delle Ricerche (IOM-CNR), c/o Dipartimento di Fisica e Geologia, Perugia I-06123, Italy}
\date{\today}

\begin{abstract}
We present a combined experimental and numerical study of the spin-wave dispersion in NiFe artificial spin-ice (ASI) system consisting of an array of  {stadium-shaped} nanoislands deposited on the top of a continuous NiFe film with non-magnetic spacer layers of varying thickness. 
 {The spin-wave dispersion, measured by wavevector resolved Brillouin light scattering spectroscopy in the Damon-Eshbach configuration, consists of a rich number of modes, with either stationary or propagating character. }
We find that the lowest frequency mode displays a bandwidth  {of approximately 0.5 GHz}, which is independent of the presence of the film underneath. On the contrary, the BLS intensity of some of the detected modes strongly depends on the presence of the extended thin-film underlayer. Micromagnetic simulations unveil the details of the dynamic coupling between ASI lattice and film underlayer. Interestingly, the ASI lattice 
facilitates {dynamics of the film} either specific wavelengths or intensity modulation peculiar to the modes of the ASI elements imprinted in the film. Our results demonstrate that propagating spin waves can be modulated at the nanometer length scale by harnessing the dynamic mode coupling in the vertical, i.e., the out-of-plane direction of suitably designed magnonic structures.

\end{abstract}

\maketitle

\section{\label{sec:intro}Introduction}

The strong coupling between nanoscale magnetic elements in two-dimensional arrays known as artificial spin ice (ASI) leads to frustrated magnetic states and results in highly degenerate energy landscapes \cite{Skjaervo}. These degenerate energy landscapes and the associated magnetic ordering in the lattice \cite{Zhou_Adv2016,Iacocca_PRB2016,Gliga_PRL2013,Jungfleisch_PRB_Ice_2016,Bang_2019,Bhat_PRB_2018,Bhat_PRB2016,Branford_PRB_2019_2,Branford_PRB_2019,Mamica_2018,Li_JPD2016,Lendinez_Nano_2021} enable the realization of reconfigurable magnonic crystals and devices. Various approaches have been proposed to utilize reconfigurability for an on-demand manipulation of the spin-wave (SW) propagation characteristics, dispersion \cite{Gliga_2020,Kaffash_PLA2021,Mont_undulation_2021,Lendinez_2019}  {and hybridizations \cite{Dion_PR_Research_2022,Gartside_2021,Lendinez_APL_2021}}. To this end, Bhat et al. demonstrated spatially localized SW modes in a connected Kagome lattice by microfocused Brillouin light scattering (BLS) \cite{Bhat_2020}, while Kaffash et al. revealed SW channels in a honeycomb array of nanodisks using microfocused BLS \cite{Kaffash_PRB2020}. Despite this progress in understanding the dynamics in ASI on the micro- and nanoscale using BLS imaging, the SW dispersion in these structures is much less explored experimentally \cite{Li_JPD2016,Li_JAP2017,Mamica_2018} and theoretically \cite{Iacocca_PRB2016,Iacocca_DMI}. Li et al. investigated the field-dependent  {spin-wave} spectra of a square ASI using wavevector-resolved BLS and revealed the soft-mode behavior in the lattice while no sizeable frequency dependence on the wavevector was found.  {Mamica et al. \cite{Mamica_2018} studied the formation of magnon band structure and gaps in anti-spin-ice systems and found that, by rotating the applied field direction, a transformation of non-propagating modes into propagating SWs with a number of hybridizations.}
Iacocca and co-workers \cite{Iacocca_DMI} theoretically explored the influence of interfacial Dzyaloshinskii-Moriya interaction and the internal magnetization configuration within an ASI vertex on the magnon dispersion. 

Furthermore, ASI structures are useful for the emerging field of \textit{vertical magnonics} that explores alternative mechanisms for controlling SWs at the nanoscale \cite{Gubbiotti_2021,Kraw_2018,Gubbiotti_book}, and the concept of ASI is already extending into three dimensions~\cite{Gliga2019MT,Fischer2020,May2019,Sahoo2021}$^,$ {\cite{Kempinger_PRL_2022,Li_APL_2022}}.
A recent theoretical proposal to modify SW channels in an ASI system is based on an ASI/magnetic thin film underlayer system \cite{Iacocca_underlayer}: while not directly demonstrated 
, the results suggest the existence of a dispersive propagating mode in the ASI-based system. However, direct proof and experimental confirmation for such a dispersive magnon characteristic in these structures remained elusive until now. 

\begin{figure}[b]
\centering \includegraphics[width=6cm]{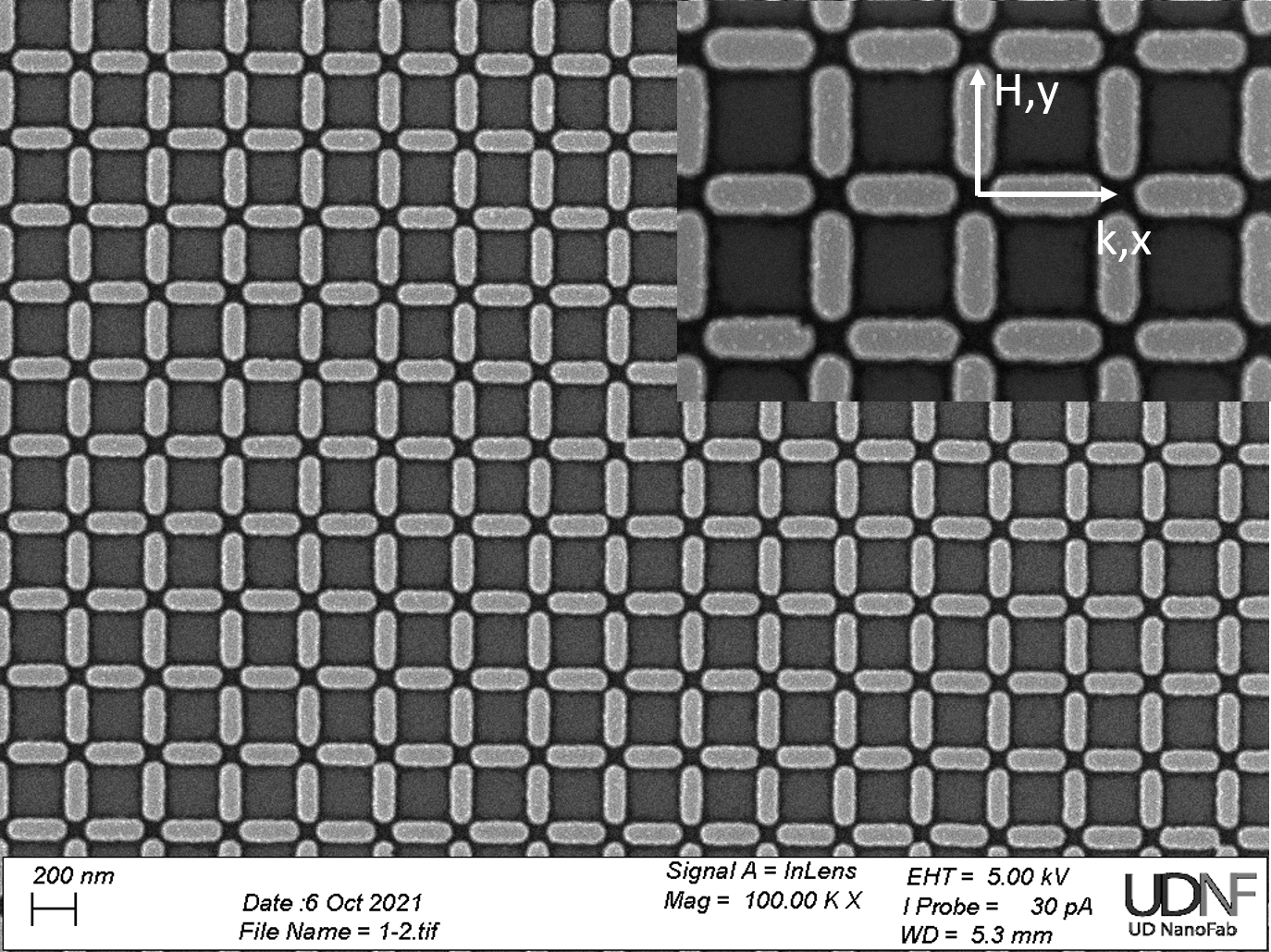}
\caption{SEM image of the artificial spin ice structure (Sample \#1). Arrows indicate the
 system of coordinate axes together with the direction of the applied magnetic field ($H$) and the probed SW wavevector ($k$). \label{fig:SEM}}
\end{figure}

Here, we couple a square ASI structure made of permalloy (NiFe)  {stadium-shaped} nanoelements (``islands'', in the following) to a continuous NiFe film underlayer to explore  {how the presence of the ASI affects the spin-wave properties in a film underlayer. }
 {We use a relatively strong magnetic field of 200 mT to ensure that the magnetic moments in a subset of islands is oriented almost parallel to the film magnetization, while the other subset is oriented in an onion-like state.}
Dependent on the separation between the underlayer and the ASI nanostructure, we observe rich SW spectra exhibiting characteristic mode dispersion crossings indicative of the dynamic coupling between the film Damon-Eshbach  {(DE)} mode and the ASI modes. Furthermore, we  {find} a dispersive magnon mode both in the isolated ASI without underlayer and in the hybrid structures. 
 {The experimental wavevector-resolved BLS spectroscopy measurements are interpreted though micromagnetic simulations, which reveal the BLS-active mode profiles, and link them to the behavior of the effective magnetic field. The existence and the significance of the dynamic coupling between the layers is also discussed as a function of the frequency and profile symmetry of the modes.}

\section{\label{sec:details}Experimental approach and micromagnetic framework}

\subsection{Sample fabrication and characterization}

We use a multi-step lithography process for sample fabrication,  {which is discussed in detail in the Supplemental Material (SM)}. The Samples consist of: (a) continuous Ni$_{81}$Fe$_{19}$ film (15~nm) (Sample \#1), (b) square ASI made of 15-nm thick Ni$_{81}$Fe$_{19}$ (Sample \#2), (c) Ni$_{81}$Fe$_{19}$ (15~nm) film with a 2~nm natural oxide spacer layer with a square ASI made of 15-nm thick Ni$_{81}$Fe$_{19}$, and (d) Ni$_{81}$Fe$_{19}$ (15~nm)/Al$_{2}$O$_{3}$ (10~nm) with a square ASI made of 15-nm thick Ni$_{81}$Fe$_{19}$. The nominal lattice constant is hence  {$a=$}$350$~nm, determining the Brillouin zone (BZ) boundary at $k=$ {$\pi/a$}$=0.9\times10^7$~rad/m. A representative scanning electron microscopy {(SEM)} image is shown in Fig.~\ref{fig:SEM} (here shown Sample \#1). As the figure shows, we achieved excellent dimension control, size, and shape homogeneity over a  {large} patterned area.

SW frequencies of the investigated samples have been measured by BLS spectroscopy from thermally excited SWs  {\cite{Carlotti_2002,Sandweg_RSI}}. 
An external magnetic field $\mu_0H=200$~mT is applied along the $y$-direction (see Fig.~\ref{fig:SEM}) and perpendicular to the incidence plane, i.e., the wavevector is in the  {DE} configuration.  {Due to the conservation of the in-plane momentum, we can detect the spin-wave wavevector by selecting the proper incidence angle of light with respect to the sample surface.\cite{Carlotti_2002,Sandweg_RSI}}. 
The magnitude of $k$ is varied from $k=0.08$ to $2.2\times10^7$ rad/m. 

\begin{figure}[bt]
\centering \includegraphics[width=0.8\columnwidth]{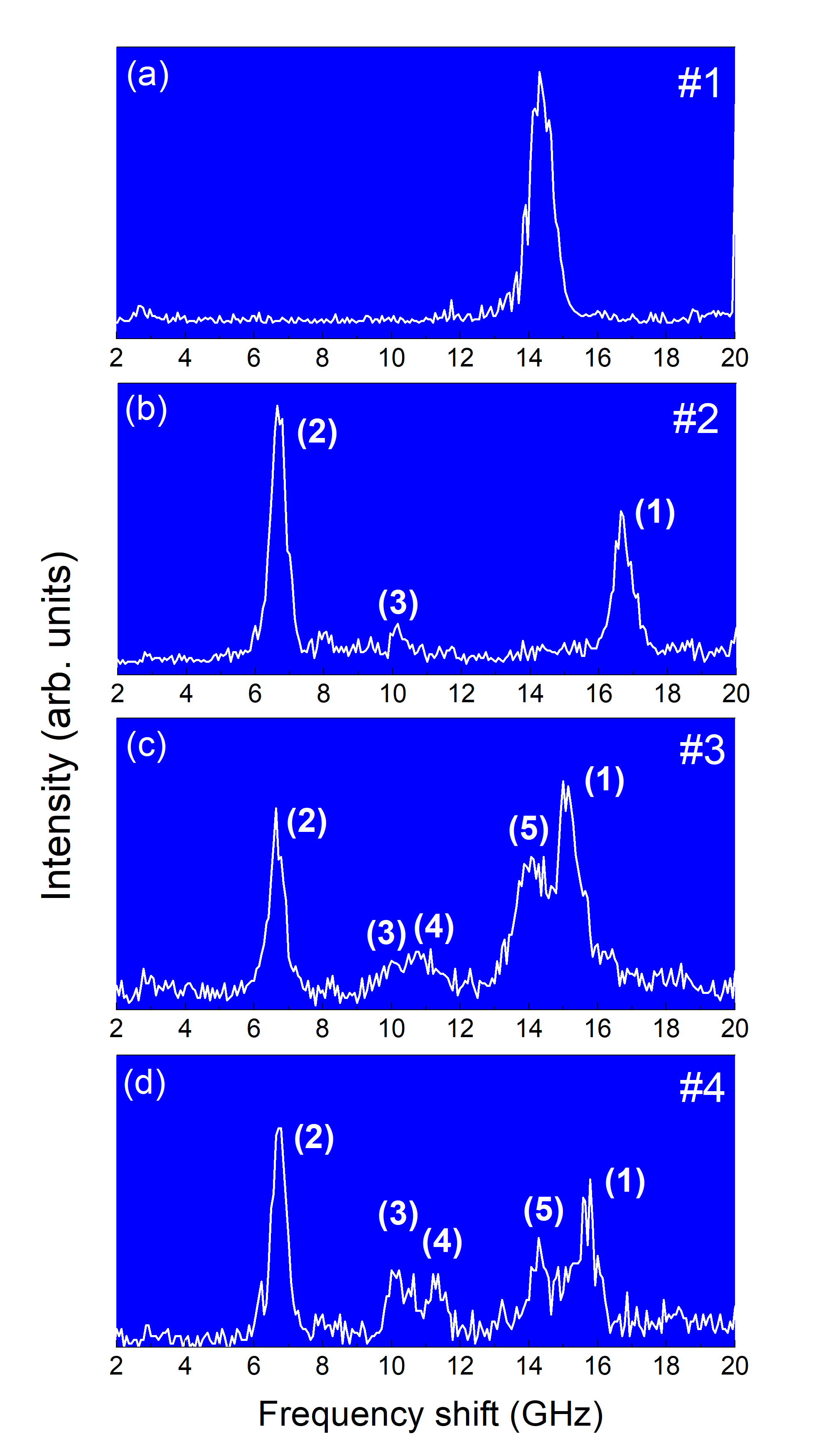}
\caption{Measured anti-Stokes BLS spectra at $\mu_0 H=200$~mT for (a) reference film (Sample \#1); (b) ASI system (Sample \#2); (c) ASI-film system with spacer 2~nm; (d) ASI-film system with spacer 10~nm. BLS peaks are labeled according to the discussion in the text.
\label{Fig:BLS_spectra}}
\end{figure}

Figure~\ref{Fig:BLS_spectra} presents a sequence of the anti-Stokes side of the measured BLS spectra at  {minimum wavevector.} 
For Sample \#1 [Fig.~\ref{Fig:BLS_spectra}(a)], only one peak is observed in the spectrum, corresponding to the  {film DE} mode.
For Sample \#2 [Fig.~\ref{Fig:BLS_spectra}(b)], two major peaks (1) and (2), and a minor one (3), are observed.
While the width of peak (2) seems to be independent of the sample, peak (1) is broader in the coupled Samples \#3 and \#4, with the presence of an additional peak, labeled as (5), visible as a left shoulder. Moreover, in the coupled samples, an additional peak (4) appears, slightly above peak (3) and with comparable intensity.

\subsection{Micromagnetic framework and simulations}
Micromagnetic simulations were performed using the graphic processing unit (GPU) accelerated software mumax$^3$ [\onlinecite{mumax3}]. The system was designed as a 15-nm-thick layer onto a 15-nm-thick primitive cell of the ASI lattice and discretized in $4.0625\times4.0625\times$5~nm$^3$ micromagnetic cells. The primitive cell is a $88\times88$ square (i.e., $357.5\times357.5$~nm$^2$) in which two islands are included, represented as ellipses\footnote{ {In simulations, we had to use the elliptical shape in order to match the experimental results: none of the many different stadium-shapes that we tried provided BLS spectra and dispersions in fair agreement with the measurements. This suggests that the “effective shape” of an island is an ellipse, although the actual shape is stadium-like as seen in the SEM image (Fig. 1). We presume that this discrepancy between experiment and simulations is due an effective (“active”) magnetic volume that is smaller than the actual island with sharper cusps at edges due to a larger oxidation at the edges.} }, each made by $64\times24$ cells, i.e., $260\times97.5$~nm$^2$.
Periodic boundary conditions are used in mumax3 in order to simulate an ideal, infinite system. The used magnetic parameters are:
saturation magnetization $M_s=800$ kA/m, exchange stiffness $A=13$ pJ/m, gyromagnetic ratio $\gamma =185\,$rad\,GHz/T. A magnetic field $\mu_0H=200~$mT is applied perpendicularly to the direction $x$ of the wavevector, with a slight tilt angle of $1^\circ$ to mimic realistic conditions. 
After relaxing the system {from uniform saturation in $y$-direction} into the equilibrium magnetic configuration, we apply a uniform sinc pulse perpendicular to the plane to excite the system, collect the excited out-of-plane magnetization and perform the time-space Fast Fourier Transform (FFT) to get either mode profiles or dispersions.  {Details on the simulation method are given in the SM.}

\begin{figure}[hp!]
\centering \includegraphics[width=7.5cm]{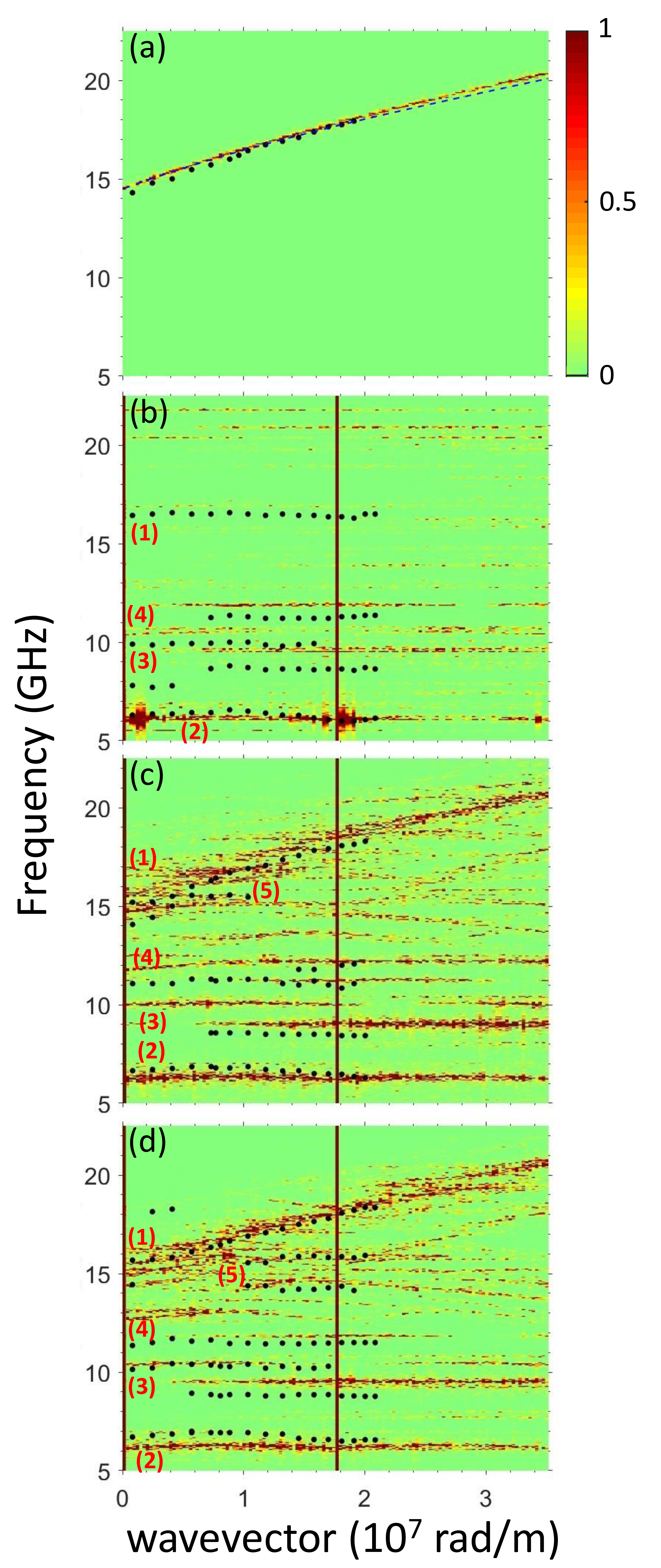}
\caption{
Measured (symbols) and simulated (color map, indicated in the side bar on the right) dispersion curves of the
(a) uncoupled reference film (Sample \#1), where the analytical dispersion is also plotted (blue dashed line) for comparison; (b) uncoupled ASI system (Sample \#2); (c) coupled ASI-film system with spacer 2~nm (Sample \#3); (d) coupled ASI-film system with spacer 10~nm (Sample \#4). The external magnetic field is $\mu_0 H=200$ mT. 
The wavevector range is extended to two BZs, and the vertical bars mark the centers $k=0$ and $k=2\pi/a\approx1.8\times10^7$~rad/m.
\label{fig:Dispersions}}
\end{figure}

\section{\label{sec:disc}Discussion}
\subsection{{\bf Discussion of the dispersions}}

The comparison between the experimental SW dispersion measured at $\mu_0H$=~200 mT (symbols) and the corresponding simulations for all investigated samples is shown in Fig.~\ref{fig:Dispersions}. 
 {Note that a large magnetic field was applied to avoid ambiguity and obtain magnetizations oriented in the direction of the applied field. This enables us to reveal the vertical coupling as a general mechanism beyond the specific ASI microstate.}

For the uncoupled film [Fig.~\ref{fig:Dispersions}(a)], the frequency dispersion is typical for the DE mode with a monotonic frequency dependence of $k$\cite{Damon,Kalinikos_1986}. For the uncoupled ASI, i.e., Sample $\#2$ [Fig.~\ref{fig:Dispersions}(b)], the measured dispersions are relatively flat (zero group velocity) except for the mode at the lowest frequency, which displays a bandwidth of about 0.5~GHz  {(see Fig.~1 in SM)}. 
Notably, the BLS intensity of some peaks is strongly dependent on $k$. This effect is reproduced  in the simulations [Fig.~\ref{fig:Dispersions}(b)]. 
For Samples $\#3$ and $\#4$ [panel (c) and (d)], a mode with monotonic dispersion is visible for frequencies higher than 14 GHz, partially overlapping with a flat dispersion visible in the low $k$-vector range at around 15 GHz (peak (5), in the figure). The measured variation of both the curves and the frequencies, compared to those of Sample $\#1$ and $\#2$, is an experimental indication of dynamic interaction between the two layers.
On the other hand, the experimental data below 14 GHz are weakly sensitive to the presence and proximity of the film underneath, which is a clear indication of the robustness of these modes with respect to the strength of ASI-film coupling.
The lowest frequency mode observed in Samples $\#3$ and $\#4$ has a frequency bandwidth similar to the uncoupled ASI system (Sample $\#2$) suggesting that the dispersion is not affected by the presence of the film underneath.

\begin{figure}[hbtp!]
\centering \includegraphics[width=1\columnwidth]{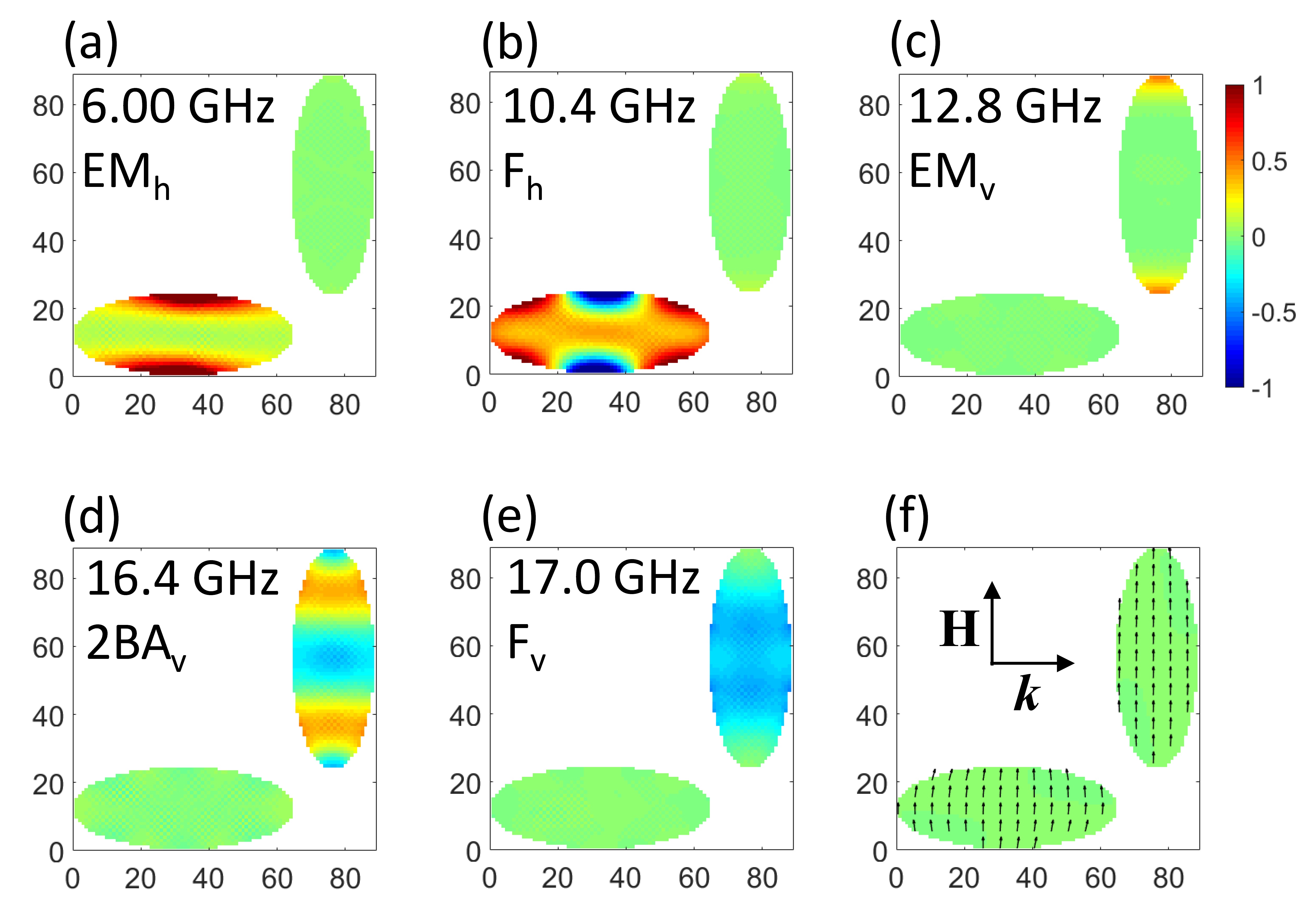}
\caption{(a-e) Phase space profiles of the main modes in the uncoupled ASI lattice, and (f) magnetization distribution, with the indication of the applied field {\bf H} 
and wavevector \textit{\textbf{k}} directions. 
Micromagnetic cell indices are shown along x and y axes. Note that mode (e) gives rise to peak (1), mode (a) to peak (2), and mode (b) to peak (3) in the spectra of Fig.~\ref{Fig:BLS_spectra}(b).
\label{Fig:FigureModesASI}}
\end{figure}

\subsection{{\bf Mode dynamics and ASI-film coupling}}\label{Modedynamics}

 {We performed micromagnetic simulations to interpret the dispersions, and unveil how ASI and film modes change and hybridize in the coupled systems.}
Hence, we can not only identify and label the peaks in the spectra but also understand the observed BLS intensity and the corresponding mode bandwidth.
In fact, the dispersion bandwidth of any mode is proportional to the dynamic stray fields it produces and hence proportional to the mode intensity (i.e., power)\cite{Zivieri_2011,Harms_2022,Gallardo_2021}.

We interpret the SW dynamics of the ASI-film hybrid system considering the layer coupling as a ``perturbation'' acting on an initial set of modes, including the DE mode of the isolated film (which at 200 mT and $k=0$ is measured at {14.3 GHz} (14.5 GHz in the simulations) and the modes of the isolated (uncoupled) ASI system [Fig. \ref{Fig:FigureModesASI}].

\begin{figure}[b]
\centering \includegraphics[width=0.8\columnwidth]{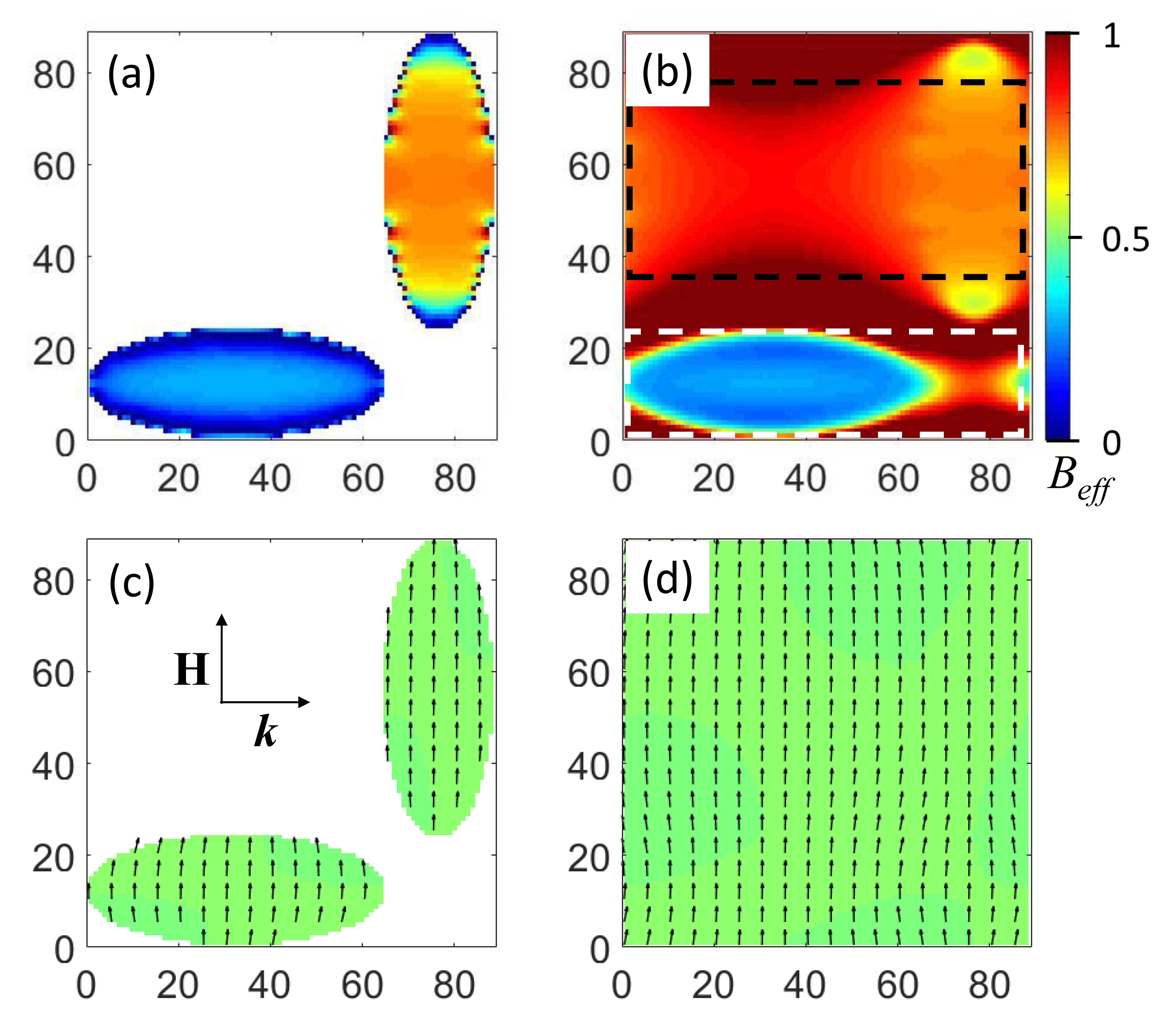}
\caption{Calculated effective internal field (${B_\mathrm{eff}}$) profiles (upper panels) and magnetization distribution (bottom panels) at 200 mT for Sample $\#3$ in (a,c) ASI layer and (b,d) film layer. 
Micromagnetic cell indices are shown along x and y axes.
In correspondence of the vertical island, in both layers ${B_\mathrm{eff}}$ has similar values. Two areas of different relative minima of ${B_\mathrm{eff}}$ are delimited by black and white dashed rectangles, separated by a high barrier (dark red areas).
\label{Fig:B_eff}}
\end{figure}

The dipolar interaction between the two layers, acting as a perturbation of the ideal uncoupled layers, has two consequences.
The first is the creation of a nonuniform magnetization in the film layer (which would be otherwise uniform), which causes a corresponding non-homogeneous ${B_\mathrm{eff}}$\cite{Bayer_APL_2003},\cite{Guslienko_2003},\cite{Bang_PRAppl2020}, that, in turns, is responsible for the non-uniform SW mode profiles. We can address this effect as a ``static coupling''. In our system, the major modification of the magnetization is found in the film layer, while the magnetization in the ASI layer is barely affected  {(compare panel (f) of Fig.~\ref{Fig:FigureModesASI} and (c) of Fig.~\ref{Fig:B_eff})}\footnote{This is due to the strong shape anisotropy of the islands.}.
The second effect is related to the dynamic dipolar field created by the precession of the magnetic moments in either layers, which we address as ``dynamic coupling''\cite{Harms_2022,Gallardo_2021}. In this case, at a given frequency, the phase profile of the mode in both layers is in some fixed relationship, the dynamic magnetization of each layer sharing some characteristics of the oscillation typical of the other layer (hybridization). 
 
\begin{figure*}[hbtp!]
\centering \includegraphics[width=2\columnwidth]{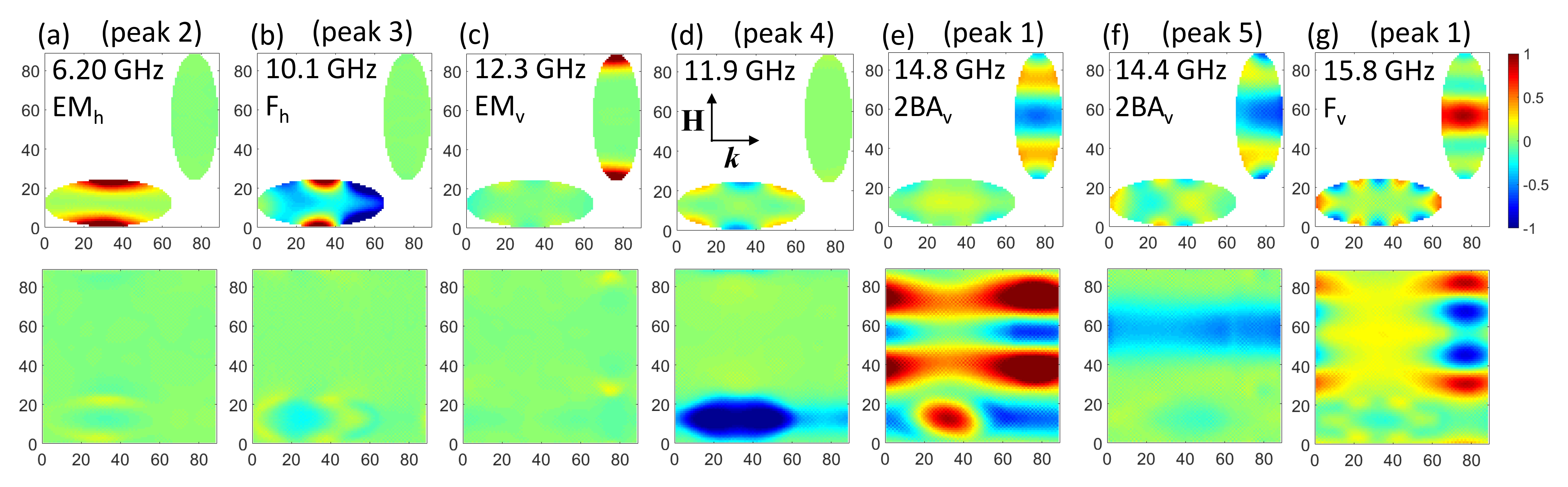}
\caption{Phase space profiles of the modes in the hybrid system Sample $\#3$, which is however illustrative also of the corresponding modes in Sample $\#4$.  {Upper (lower) panels show dynamic magnetization in the ASI (film) layers.}
Micromagnetic cell indices are shown along x and y axes. 
Modes (a-c) show no hybridization between ASI and film layer and look like those in Fig.~\ref{Fig:FigureModesASI}(a-c). Mode (d) is very intense only in the film layer, where it is reminiscent of the film DE mode. This mode is trapped in the deep minimum of ${B_\mathrm{eff}}$ [compare to Fig.~\ref{Fig:B_eff}(b)]. In panel (d) we also show the field and wavevector directions as a reference. Modes (e-g) are the result of the dynamic coupling between the former DE mode of the film and either the ASI mode $2BA_v$ (e,f) or $F_v$ (g). 
The peak numbers refer to the measured spectra.
\label{Fig:FigureModes}}
\end{figure*}

 {Using micromagnetic simulations, we find that modes of the hybrid systems with a frequency below the DE-film frequency have the same frequency and spatial profiles as the isolated ASI. Furthermore the ASI layer of the hybrid system exhibits the same space phase profile, while their film underlayers show negligible dynamic magnetization. Thus, the two layers are dynamically uncoupled in this frequency range (compare Fig.~\ref{Fig:FigureModes} with Fig.~\ref{Fig:FigureModesASI}). This observation is independent of the spacer layer thickness.} 
 {On the other hand, the modes detected by BLS above 14~GHz are hybridized, which means that at a given frequency the two layers show similar dynamic magnetization profiles that are correlated. This effect is robust with respect to {variations of the} spacer layer thickness (and the applied magnetic field). We refer to this type of coupling as \textit{dynamic coupling}.}
 {Previous works reported various types of mode hybridizations\cite{Gubb_ell_2005,Vysotskii_2008,Lendinez_2019,Gartside_2021,Dion_PR_Research_2022} and  conversion\cite{Mont_vort2012,Tacchi_2012}. {However, hybridization in these cases occurred among modes relative to the same system geometry, and hence the
same underlying magnetization. }
 {In our case, the hybridization is distinctly different from the previous reported ones: here,  the hybridization occurs between modes belonging to different layers.} {i.e., physical structures with different geometry (film and ASI) and different underlying magnetization configuration
[compare Figs.~\ref{Fig:B_eff}(c) and (d)].}
 

 {Micromagnetic simulations reveal a variety of hybrid modes. Here, we illustrate the modes which likely have a sufficiently strong intensity to be detectable experimentally.}
In the following, we label ASI modes residing in the vertical islands with subscript ``v'' and those in the horizontal islands with subscript ``h''.

 {We describe the experimentally obtained dispersions based on our simulation results.}
First, the edge mode profile of the horizontal island $EM_h$ is shown in Fig.~\ref{Fig:FigureModesASI}(a). Although $EM_h$ is mainly localized at the edges along the field direction, the broad area of the oscillation produces large dynamic dipolar fields responsible for the previously discussed bandwidth\cite{Mont_Asymmetry2013}.
The corresponding edge mode of the vertical island $EM_v$ occurs at $\approx$ 12.8~GHz, Fig.~\ref{Fig:FigureModesASI}(c).  {We suspect this mode does not lead to a detectable BLS peak since it is localized in a narrow region.}
The fundamental mode of the horizontal island $F_h$ is calculated at 10.4~GHz [Fig.~ \ref{Fig:FigureModesASI}(b)], and is hybridized with an edge mode: this reduces its average intensity, appearing in the BLS spectra as peak (3) at about 10~GHz. Its calculated dispersion is rather flat, in agreement with the measured one.
The fundamental mode of the vertical island $F_v$ appears at 17.0 GHz [Fig. \ref{Fig:FigureModes}(e)], shows no nodal lines, and hence has a large intensity that can be associated to peak (1) of the BLS spectra measured at 16.5 GHz. In spite of the large intensity, the dispersion appears flat as well. 
We also show the profile of an additional mode of the vertical islands at 16.4 GHz exhibiting two nodal lines perpendicular to the magnetization direction $2BA_v$ [BA stands for backward-volume character, and 2 refers to the node number, Fig.~\ref{Fig:FigureModesASI}(d)].

When the film underlayer and ASI layer interact, we observe two main effects: 1) a change of symmetry (and hence amplitude) of the modes of either layer since their static magnetization changes and 2) an ASI/film hybridization of the modes, which arises if the corresponding resonance frequencies are close. 
Two out of the four ASI-film hybridized modes [Fig.~\ref{Fig:FigureModes}(d,f)] do not have any nodal lines in the film layer and, hence, can be seen as fundamental modes with a strong localization in the two different relative minima of ${B_\mathrm{eff}}$ [Fig.~\ref{Fig:B_eff}(b)], the deepest one occurring below the horizontal island, where the magnetization is less uniform. The other two show a phase oscillation with nodes within the primitive cell [Fig.~\ref{Fig:FigureModes}(e,g)]. 
 
The first of these four modes [Fig. \ref{Fig:FigureModes}(d)] consists of a high order edge mode in the ASI layer (of negligible intensity) and an intense oscillation in the film layer, which is reminiscent of the former DE mode of the film, now strongly localized within the absolute minimum of ${B_\mathrm{eff}}$, hence at a much lower frequency. ${B_\mathrm{eff}}$ increases with increasing spacer, and so does the frequency of this mode: 11.9~GHz for Sample $\#3$ and 12.7~GHz for Sample $\#4$.
The intensity of this mode is rather large and emerges in the spectra as ``new'' peak (4), absent in the uncoupled systems. Simulations show a weakly increasing dispersion at low $k$, which is absent in the experimental results.
 {We observe that its mode profile in the ASI layer changes if the separation changes: }
 {this means that there is no dynamic coupling and the effect is purely coincidental.}

 {A general rule can be derived from our simulation results: if for any given mode the dynamic coupling is present, the  dynamic magnetization profile in either of the two layers is mutually hybridized, i.e., linked by a common precise dynamic quantity (wavelength, phase coherence, symmetry, etc.). This effect is robust and independent of the spacer layer and the applied magnetic field. Conversely, if the dynamic coupling is absent, the associated ASI layer and the underlayer may show  different profile when an external parameter is changed.}
 
The second fundamental mode derived from the film DE [Fig. \ref{Fig:FigureModes}(f)] consists of a $2BA_v$ mode in the ASI layer and a quasi uniform oscillation in the film layer, which is again reminiscent of the DE of the uncoupled film, though weaker and localized in a narrow area inside the black rectangle of Fig.~\ref{Fig:B_eff}(b)]. Its frequency is higher than for mode (d) [14.4 GHz both for Sample\#3 and \#4, peak (5)] because the ${B_\mathrm{eff}}$ value it experiences is higher. 
This time, this mode is indeed the result of the dynamic coupling between ASI and film layers, because of the coherence and vertical correspondence of the oscillation profiles in the layers. We verified that, differently from mode (d), in this mode the relationship between the layers is persistent independently of any parameter change.

The dynamic coupling is particularly evident for the mode of Fig. \ref{Fig:FigureModes}(e), with a perfect vertical correspondence of the in-phase precession of the two layers.  {The mode consists of a $2BA_v$ mode in the ASI layer, while the oscillation in the film layer resembles a backward SW  with $\lambda=a/2$. This wavelength is imprinted by the ASI oscillation onto the film SW. In simulations this mode produces an intense peak at 14.8~GHz. Hence, we associate this mode (e) to peak (1) in the coupled systems.}


 {Finally, based on the simulations, we find that the mode shown in Fig.~\ref{Fig:FigureModes}(g) at 15.8 GHz for Sample \#3 (15.5 GHz for Sample \#4) exhibits a $F_v$ profile in the ASI layer and a quasi-uniform oscillation in the film layer. However, a remarkable out-of-phase precession of dynamic magnetization is found for the two layers in vertical direction. These features are independent of the spacer and the applied field. This implies this mode arises  from the dynamic coupling. The intensity of this mode is found to be very large. Hence, we associate this mode to peak (1) [panel (e)] leading to a broadening of the peak.} 

 {The three modes (e-g) in the hybrid system are responsible for the monotonic dispersive curve starting from 14~GHz resembling  the DE dispersion of the uncoupled film. With the present resolution (both experimental and computational) we cannot conclusively determine whether the structure of the dispersion at low $k$ between 14 and 16~GHz corresponds to a crossing or an anticrossing of modes (compare to Ref.~[\onlinecite{Kraw_2018}]).}

Other modes (often with incompatible profile symmetry or different node numbers) might still ``occasionally'' appear at the same frequency in the associated ASI and film layers. However, the reason is, in this case, pure coincidence, not interaction; 
hence, they are not dynamically coupled.

The dynamic coupling  {between layers of different geometry that was} unveiled here is particularly interesting since it demonstrates the possibility of imprinting a specific wavelength, determined by design of the ASI element and the lattice constant of the ASI to the film  [$\lambda=a/2$ for mode Fig. \ref{Fig:FigureModes}(e)]. This effect is three-dimensional since the vertical coupling determines the wavelength of the in-plane propagation. 

\section{\label{sec:summary}Conclusions}
In conclusion, with studied experimentally and numerically the spin-wave dispersion in dipolarly coupled hybrid ASI/film system separated by a non-magnetic spacer. 
The experiments, supported by the micromagnetic simulations, reveal that the coupling between the layers arises in correspondence to the film mode frequency. Below that frequency only the ASI layer is excited and exhibits mode profiles and frequencies almost identical to the uncoupled ASI lattice.
Analysis both of the dispersions and the BLS spectra highlighted an appreciable bandwidth of about 0.5~GHz for the lowest frequency mode, almost insensitive to the ASI/film coupling and spacer thickness. The dispersion of the DE mode in the uncoupled film is instead found to be due to hybridized ASI/film modes, which showed interesting properties derived from the geometric design of the layers.
Finally, we demonstrated the possibility of creating spin waves with particular wavelengths or dynamic intensity modulations in the continuous film by suitably designing the ASI geometry of the overlayer  {, thus adding a new degree of freedom to control the spin-wave propagation in vertical magnonic systems.}


\section{\label{sec:SM}Supplementary Material}
See supplementary material for details on the experimental approach, micromagnetic framework, and the experimental bandwidth estimation of the ASI mode $EM_h$.

\begin{acknowledgments}
Research at the University of Delaware, including sample design and fabrication, and initial micromagnetic simulations supported by the U.S. Department of Energy, Office of Basic Energy Sciences, Division of Materials Sciences and Engineering under Award DE-SC0020308. G.G. acknowledges financial support from the Italian Ministry of University and Research through the PRIN-2020 project entitled “The Italian factory of micromagnetic modelling and spintronics”, cod. 2020LWPKH7.  {This project has received funding from the European Union’s Horizon 2020 research and innovation programme under grant agreement No 101007417 having benefited from the IOM-CNR access provider in Perugia access site within the framework of the NFFA-Europe Pilot Transnational Access Activity, proposal ID151.}
\end{acknowledgments}
\bibliography{aipsamp}

\end{document}